# Timing Constraints Support on Petri-Net Model for Healthcare System Design

Sabri Mtibaa and Moncef Tagina

**Abstract**— The worldwide healthcare organizations are facing a number of daunting challenges forcing systems to benefit from modern technologies and telecom capabilities. Hence, systems evolution through extension of the existing information technology infrastructure becomes one of the most challenging aspects of healthcare. In this paper, we present a newly architecture for evolving healthcare systems towards a service-oriented architecture. Since healthcare process exists in temporal context, timing constraints satisfiability verification techniques are growing to enable designers to test and repair design errors. Thanks to Hierarchical Timed Predicate Petri-Net based conceptual framework, desirable properties such as deadlock free and safe as well as timing constraints satisfiability can be easily checked by designer.

**Index Terms**— Healthcare; information technology; service-oriented architecture; Hierarchical Timed Predicate Petri-Net; conceptual framework; timing constraints satisfiability.

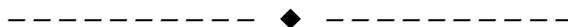

## 1 INTRODUCTION

The healthcare industry is diverse, complex and experiencing many challenges regardless of region, country or system. In fact, healthcare systems around the world are being impacted in many different ways: providers and payers are facing increasing pressure to deliver higher quality of care and service whilst at the same time healthcare costs are increasing drastically and funding is being reduced [1].

What's more, the lack of effective information exchange between a healthcare organization's facilities, departments and staff – or the means for quicker and more convenient patent/physician consultations and discussions – reduces the ability to deliver patient services and treatment.

In the face of all these challenges, there is one overall reality: a key element of the healthcare industry's ability to thrive and succeed into the future rests on the strengthening and assurances of effective communication and collaboration between patients and healthcare providers.

The focus on the climbing cost to deliver and maintain quality healthcare is no longer in the peripheral view, but a clear line of sight for the patients, healthcare providers, regulators, and payers. This brings new requirements for healthcare professionals to share information, communicate and collaborate in real time from multiple locations, because medicine is a collaborative science. Communications become a strategic asset for a strongly needed healthcare transformation technology. It must be deployed to this field in order to ensure better context for medical decisions, reduce administrative costs and improve patient safety by reducing errors. In our previous work [2], a distributed telemedicine environment was exposed to accelerate service innovation for personalized and blended medicine services. In this paper, our efforts concentrate on the modeling of healthcare system to allow properties analysis and verification at the design before deployment. This model will allow designers to detect erroneous properties and formally verify whether the service process design does have certain desired properties.

Nowadays, the need to transform from the current hospital centralized treatment-based mode to prevention-oriented comprehensive healthcare mode in which hospitals, communities, families and individuals are closely involved become on the top of professional's agenda. The new mode needs to provide individuals with intelligent health information management and healthcare services with a special attention to timing constraints (notification in case of urgency, remote session enablement …). This work is the extension of the distributed telemedicine environment to enable monitoring of patients outside hospitals or medical centers thanks to telecommunication technologies.

The advancement of information technology (IT) brings more opportunities for innovations in the healthcare area [3]. The use of service oriented technologies such as SOA, Web Services allows service providers to reduce and simplify integration process, to abstract network capabilities (e.g., call control, presence, location, etc.), and create personalized and blended services (both internally and with 3rd party partners). These technologies facilitate the construction of service systems with higher reusability, flexibility, extensibility, and robustness.

Cloud computing is evolving as an important IT service platform with its benefits of cost effectiveness and global access. To become a widely adopted IT infrastructure and service platform, cloud computing has to be integrated

————————————————

- *Sabri Mtibaa is with the LI3 Laboratory, National School of Computer Sciences, Manouba, 2010. E-mail: Sabri.Mtibaa@gmail.com.*
- *Moncef Tagina is with the LI3 Laboratory, National School of Computer Sciences, Manouba, 2010. E-mail: Moncef.Tagina@ensi.rnu.tn.*

with other systems in organizations. In academia, there is very limited study of cloud computing integration. In practice, the industry lacks comprehensive systems integration architecture or tools that can integrate any system universally [3]. Built upon Enterprise Service Bus (ESB) as an integration backbone, this paper presents a novel citizen-centric healthcare service platform. We propose also a graphical conceptual framework which helps to design healthcare system based on a traditional healthcare scenario. Since healthcare service is critical, timing constraints specification must be considered. So the concepts of healthcare process net and hierarchical web service net under time constraints specification are defined by Hierarchical Timed Predicate Petri-Nets (HTPPN).

The remainder of this paper is organized as follows. In Section 2, we present the healthcare service platform architecture. A scenario using the novel healthcare is introduced in Section 3. In Section 4, we describe our conceptual framework for healthcare system design. Section 5 provides the model of the conceptual framework using Hierarchical Timed Predicate Petri nets. Finally, we conclude in section 6.

## 2 HEALTHCARE SERVICE PLATFORM ARCHITECTURE

In this section, we present the global context of our work and an overview about service oriented architecture, cloud computing and enterprise service bus. Then, the healthcare services platform architecture is exposed and some basic concepts and definitions are explained.

### 2.1 Concepts

1) *Service Oriented Architecture (SOA):*

In this IT architecture, applications and more discrete software functions are network-based, loosely coupled and available on demand to authorized users or to other applications or services. Although SOA is not a new concept, the emergence of Web services as a standard way to expose, describe, access and combine services has given new life to this approach to computing. The key idea of SOA is the following: a service provider publishes services in a service registry [2]. The service requester searches for a service in the registry. He finds one or more by browsing or querying the registry. The service requester uses the service description to bind service. These ideas are shown in Fig. 1.

2) *Cloud Computing:*

Cloud computing called also *utility computing* refers to an IIT service model and platform that provides on-demand based IT services over the internet (see Fig. 2). The five essential characteristics are: on-demand self-service, broad network access, resource pooling, rapid elasticity, and measured service [3]. The three services models include:

- SaaS (Software as a Service) which delivers software service on demand, such as, CRM service and Google Gmail;

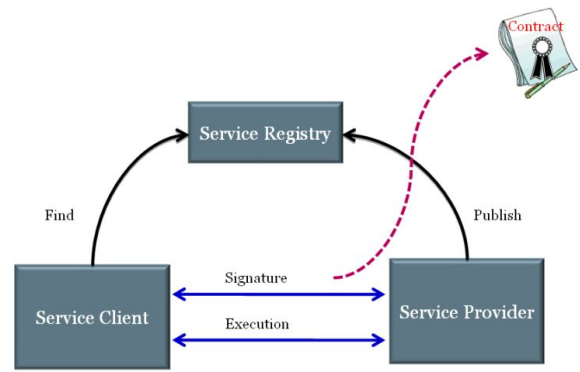

Fig. 1. Reference architecture of web services- SOA

- PaaS (Platform as a Service) which provides the computing platform for companies to deploy and customize business applications on demand, such as, Google App Engine and Microsoft's Azure;
- IaaS (Infrastructure as a Service) which offers data center, infrastructure hardware and software resources on demand, such as, Amazon Elastic Compute Cloud (EC2) and VMware vCloud Datacenter. Both of these resources provide virtual computers for renters to run their business applications.

The four major deployment models include: private cloud, public cloud, community cloud, and hybrid cloud.

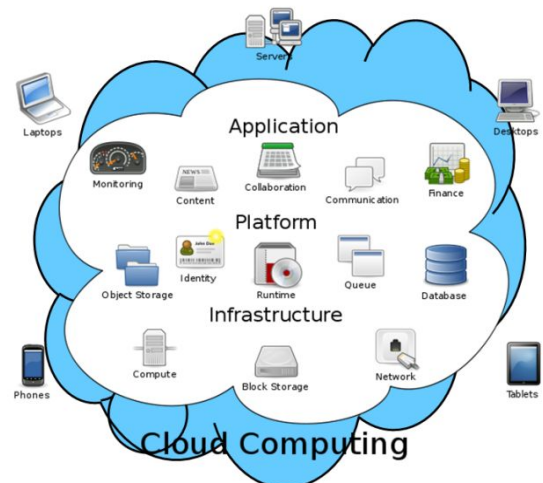

Fig. 2. Cloud computing architecture

Companies normally adopt different service models and deployment models depending on their unique business processes and demands on IT services.

Cloud computing today is an evolution and application of modern ICT including server virtualization, autonomic computing, grid computing, server farm, network storage, and web service.

3) *Enterprise Service Bus:*

ESB is one piece of an infrastructure that might help facilitating the implementation of a SOA, but it is not a perquisite. There are many aspects of an ESB that fit well with the SOA model, and denying its possible usefulness would be counterproductive, but the two are not completely inter-dependent [4].

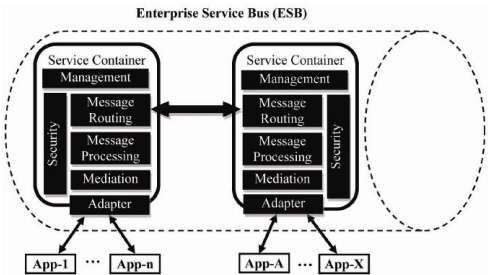

Fig. 3. ESB architecture

Fig. 3. depicts the base functional elements within an ESB. It includes:

- Data transformation.
- Application adapters.
- Automation of processes.
- Transformation.
- Routing.
- Messaging.
- Event triggering.

If we consider some of these functional elements it can be seen that items such as application adapters fall neatly into the product category, while routing and messaging are more of an architectural consideration.

## 2.2 System Architecture

First, we present our Healthcare Service Platform (HSP). It intends to provide personalized healthcare services for the public. The healthcare value chain is complex. It consists not only of healthcare providers, but also of payers (government, employers and patients), fiscal intermediaries, distributors and producers of pharmaceuticals and devices. The HSP does not attempt to address this complete value chain. It focuses on the delivery of healthcare services. It is an end-to-end reference architecture that focuses on meeting the needs of citizens, patients and professionals. Its architectural diagram is given in Fig. 4.

We distinguish three main components, i.e. body sensor networks (BSN), IaaS cloud, healthcare delivery environment.

- BSN: according to circumstances and personalized needs, appropriate health information collection terminals (i.e. sensors) are configured for different individuals. BSN is used to provide long term and continuous monitoring of patients under their natural physiological states. It performs the multi-mode acquisition, integration and real-time transmission of personal heath information anywhere [5].
- IaaS cloud: modern healthcare is information driven. Healthcare providers are making progress in building an integrated profile of patients. This data sits in systems throughout the enterprise including the HER and many other electronic systems throughout the enterprise and community [6]. This component achieves the rapid storage, management, retrieval, and analysis of massive heath data. It mainly includes *Electronic Medical Record* (EMR) repository. It considers also personal health data acquired from BSN.
- Healthcare delivery environment: it includes a personal health information management system. It replaces expensive in-patient acute care with preventative, chronic care, offers disease management and remote patient monitoring and ensures health education/wellness programs.

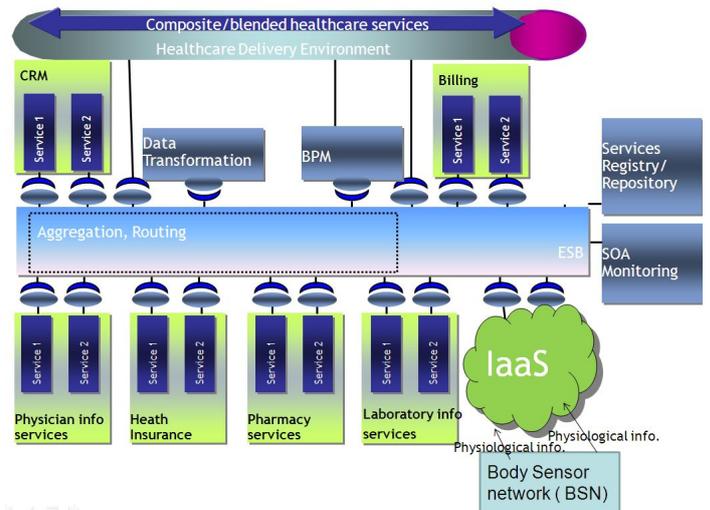

Fig. 4. HSP architecture

## 2.3 Healthcare Web Services Provided by HSP

In PHISP, we adopt the design idea of SOA and Web service technology for its design and implementation. The majority of its functional modules are developed and packaged in the form of services. Here, we overview some of them as follows.

- *PhysInfoWS*: this service can acquire some general physiological signals such as body temperature, blood pressure, and saturation of blood oxygen, electrocardiogram, and some special physiological signals according to different sensor deployment for different users. User's ID number is required.

- *EnvInfoWS*: for a unique ID number, this service can acquire temperature, humidity, air pressure and other environmental information for this user.
- *SubjFeelWS:* it can acquire the user subjective feelings, food intake, etc., and the information is often provided by the user from the terminal.
- *CoronaryDiagWS:* it can analyze the information according to a series of analysis models, which are built for coronary heart disease, and then produce preliminary diagnostic results.
- *AssessmentWS:* this service can assess the status of the patient's health risk based on the diagnostic results and the EMR information of the patient.
- *EmrWS:* this service can output the user's medical history information.
- *GeoWS*: it can return the user's location.
- *EmerWS*: it can raise an alarm to the user in case of illness.
- *GuideWS*: it can provide the patient with preventive measures especially items that need attention.

## 3 HEALTHCARE SERVICE SCENARIO

A way to motivate and illustrate this work, we presents an example of healthcare service scenario. We distinguish three main layers: service, business and HSP. The service layer consists of available web services, and the business layer represents the Web service like operations typically ordered in a particular application domain. We refer to the selected services as member services (see Fig. 5).

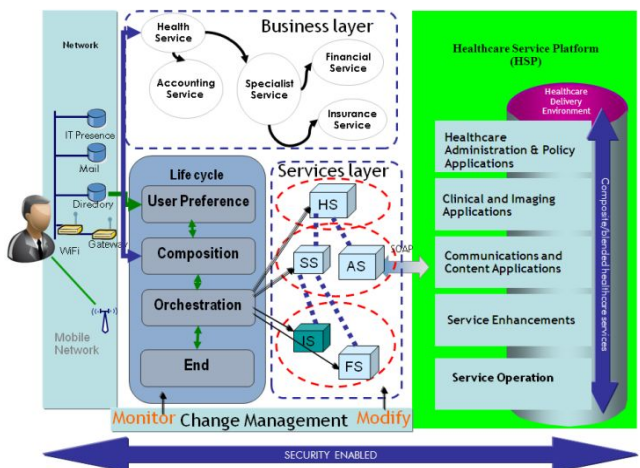

Fig. 5. Healthcare Service Scenario based on HSP

Key healthcare environment objectives include:

- **Allowing people to stay in their homes to an older age.** By doing this, we can reduce the economic burden of dedicated care facilities and improve quality of life for a substantial proportion of the aging population.
- **Using televisions to keep in touch.** Another use of camera technology is in conjunction with an IPTV set-top box and connection back to a Contact Center.

Let us assume that a *citizen* establishes a need for a business objective (healthcare service). Typically, he starts with formulating the business strategy (or goal). During the planning, some services can be identified: *HealthService*, *AccountingService*, *SpecialistService*, *FinancialServcie* and *InsuranceService*. Second, the *senior citizen* develops a specification listing the services to be composed through a graphical interface. We assume that *HS*, *AS*, *SS*, *FS* and *IS* are selected and orchestrated. The third step is the orchestration where member services that match the specified high level configuration are selected and invoked. We describe here the ideal scenario: the *senior citizen* subscribe to *HealthService*. Then all information regarding who contacts it and when are forwarded to *AccountingService*. *HealthService* forwards also the received data to *SpecialistService* in charge of checking the received values. After analyzing the received values, the team sends a confirmation or an adjustment of the medication doses. The *FinancialService* and *InsuranceService* are executed to finalize the process.

## 4 CONCEPTUAL FRAMEWORK FOR HEALTHCARE SYSTEM DESIGN

In order to design the healthcare service scenario, we represent the three layers in hierarchical structure: customization layer, business layer and execution layer (Fig. 6).

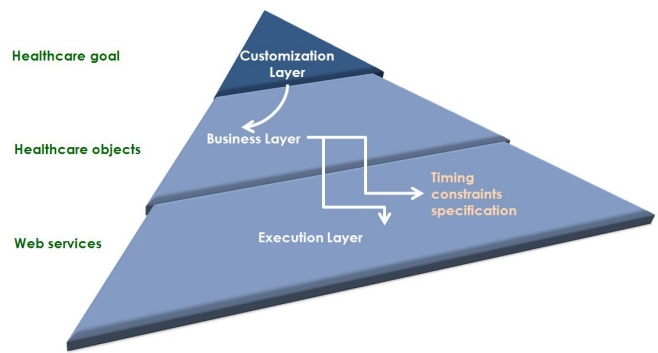

Fig. 6. Conceptual framework of healthcare system

- *Customization Layer*

This layer aims to adapt efficiently the system model to citizen needs.

- *Healthcare process layer*

Which includes a set of healthcare objects is designed based on instructional concepts such as activities and services to patients. This means that the layer reflects the pure healthcare perspective.

- *Execution layer*

Which integrates web services is designed upon the technological concepts such as web service or web service composition. It reflects the technology perspective. A variety of features and components can be realized as web services. A web service can be either simple or complex. Web services are mapped healthcare objects from healthcare process layer to web services layer in simple or complex healthcare scenarios.

When using layered architecture, the design must adhere to the following design requirements:

- Aggregation - n healthcare objects (HO) 'a' in the healthcare process layer corresponds to a set of web service ($a_1, a_2… a_n$) in the execution process layer, where those activities jointly exchange the same messages as the HO 'a', thus achieving the same goal.
- Condition alteration – conditions must be designed such that the HO preconditions are the same or weaker than '$a_1$' in the execution process layer. HO postcondition must be the same or stronger than '$a_n$' in the execution process layer.

## 5 MODELING THE CONCEPTUAL FRAMEWORK BY HIERARCHICAL TIMED PREDICATE PETRI-NETS

Petri-Nets are well-established process-modeling approach. Petri-Net represents communication patterns, control patterns, and information flows. They are used mainly for modeling and analyzing with formal semantics, powerful expressiveness and abundant analysis methods.

As Petri-nets find their way into different research and application areas, there are many extensions of the original Petri net definition. Hierarchical Petri-Nets were developed by Valette [7]. In this paper, an assumption is made that the healthcare process description represents a well-formed workflow process which respect the time constraints (date intervals). So, we will be able to transform the operational semantics of conceptual framework discussed above into Hierarchical Timed Predicate Petri-Nets (as described in TABLE I).

TABLE I. HTPPN-CONCEPTUAL FRAMEWORK MAPPING

| Semantic Map | |
|---|---|
| HTPPN | Conceptual framework |
| Place | condition of an action (pre-condition/post-action) |
| Transition | Action |
| Token | Process state |
| Arc | Process action flow |

- *Definition: Healthcare process Net*

The algebraic structure of Healthcare Process Net named HPN= (P, T, TR, F, $P_i$, $P_0$) if the following conditions hold:

- $T^R \subseteq T$
- $F \subseteq (P \times T) \cup (T \times P)$
- $P_i \in P$
- $P_0 \in P$

P and T are the sets of places and transitions, respectively, representing the pre/post conditions and implicit/explicit action in healthcare process.
$T^R$ is a finite set of refinable transitions. And R a function that associates with every refinable transition $t \in T^R$ an Execution process Net.
F is a finite set of arcs representing healthcare process action flow.
HPN is Healthcare Objects structured Petri Net which describes the structure and manual dependence of set healthcare objects. It allows modeling the context of each healthcare object in terms of preconditions (prerequisites) and postconditions (HO).

- *Definition: Execution process Net*

The algebraic structure of Execution process Net named EPN = (P, T, F, N, TC, TD, $P_i$, $P_0$) if the following conditions hold:

- $F \subseteq (P \times T) \cup (T \times P)$
- N: T → Name ∪ {$\phi$}, Name represents the name of web services composition. $\phi$ represents that the operation of web service is not actual execution but structural simulation.
- $P \cap T = \varnothing$
- $P \cup T \neq \varnothing$
- $P_i \in P$
- $P_0 \in P$

P is a finite set of places representing the state of web service. T is a finite set of transitions which represents the activity of web service. F is called the web services action flow. TC = TC(p)∨TC(t), where TC(p) is a set of all place time pairs and TC(p)={ [$TC_{min}(p), TC_{max}(p)$]∈Z × Z | $TC_{min}(p) < TC_{max}(p) \wedge p \in P$ }, TC(t) is a set of all transitions time pairs and TC(t)={ [$TC_{min}(t), TC_{max}(t)$]∈ Z × Z | $TC_{min}(t) < TC_{max}(t) \wedge t \in T$ }. TD∈Z, is a set of time duration.
The interval [$TC_{min}(p), TC_{max}(p)$] denotes the time period during which $p_i$'s succeeding transitions are enabled after a token arrives at a place $p_i$. $TK_{arr}(p_i)$ and $TK_{lea}(p_j)$ are the time at which a token arrives and leaves at a place pi.
For t ∈ T, if TC (t) ∈ [$TC_{min}(t), TC_{max}(t)$], then a transition t can fire at least after $TC_{min}$ unit time intervals if it is enabled at marking M; during the period, transition t must fire at most after $TC_{max}(t)$ unit time intervals if there is no transition enable, which may change marking and make the transition t unreachable. The firing duration TD(t) which corresponds the execution duration of a transition, denotes how long the firing of t will last.

In order to represent timing constraints, we need to augment the execution model with the following basic temporal types [8]: time point, duration and interval constraints. They are defined as follows:

TEB(j) is the sum of Message Delay Time and waiting time of the activity $j$ of a process instance. [$TEB_{min}(j), TEB_{max}(j)$] denotes the time period during which the activity j are enabled after the intermediate preceding activity of the activity $j$ finishes.

[$TEC_{min}(j), TEC_{max}(j)$] denotes the time period during which the activity j can be executed after it is enabled.

A web service behavior is basically or partially ordered set of activities.

In order to construct a more complex web services net, the designer will be able to construct a more complex web services nets using some elementary Petri-Nets.

We figure out some useful patterns defined in Workflow Management Coalition (WFMC):

- *Sequential structure :*

The sequence construct allows the designer to define a collection of activities to be performed sequentially in lexical order (see Fig. 7.a). A sequence activity contains one or more activities that are performed sequentially.

The processes *i, j* execute in sequential order, and:
TEB(i o j)= TEB(i)+ TEB(j)+ TEC($p_i$)

- *Parallel structure :*

The flow construct allows the designer to two or more activities to be executed in parallel, giving rise to multiple threads of control. Transition $T_0$ (see Fig. 7.b) is a point where a single thread of control splits into two or more threads that are executed in parallel, allowing multiple activities to be executed simultaneously.

The processes *i, j* execute concurrently:
TEB(i || j) = Max(TEB(i),TEB(j))

- *Conditional structure:*

The flow construct allows the designer to select exactly one branch of execution from a set of choices. It supports conditional routing between activities. Place P*i* (see Fig. 7.c) is a point where a single thread of control makes a decision upon which branch to take when encountered with multiple alternative activity branches.

Suppose process *i* as preceding process, and process *j, k* execute alternatively:

TEB(i o (j V k)) = Max(TEB(i) + TEB(j) + TEC($p_j$), TEB(i) + TEB(k) + TEC($p_j$)

- *Loop structure :*

The loop construct (Fig. 7.d) allows the designer to indicate that an activity is to be repeated until a certain success criteria has been met. A while activity supports repeated performance of an activity in a structured loop, that is, a loop with one entry and one exit point.

If we suppose that process *i* execute k times,

TEB(k x i) = ($\Sigma_{j=1,k}$ $TEB_{min}(j)$, $\Sigma_{j=1,k}$ $TEB_{max}(j)$)

- *Application: Healthcare scenario*

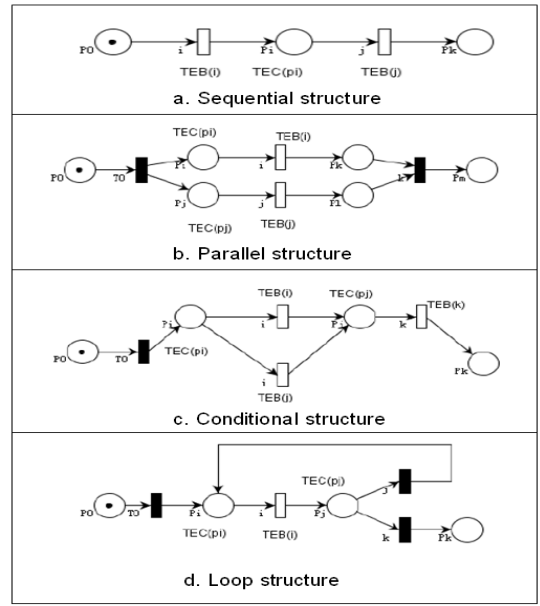

Fig. 7. Patterns defined in Workflow Management Coalition

From the above, the first part of hierarchical healthcare process net N with refinable transition named 'Health Service is shown in Fig. 8.

The planned web service is the assembly of the set of web services presented previously (*PhysInfoWS, EnvInfoWS, SubjFeelWS, CoronaryDiagWS, AssessmentWS, EmrWS, GeoWS, EmerWS, GuideWS*).

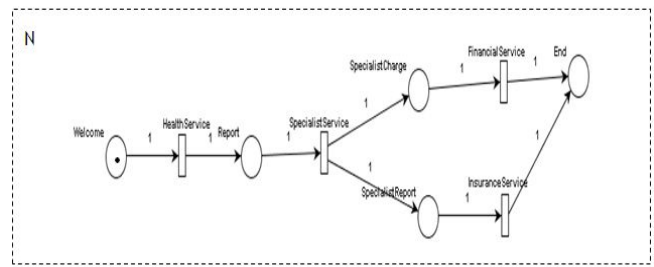

Fig. 8. A healthcare process modeled (High level)

HealthService transition is refined with the attachment of web services net N' (see Fig. 9). We added some timing constraints to different web services and operations involved.

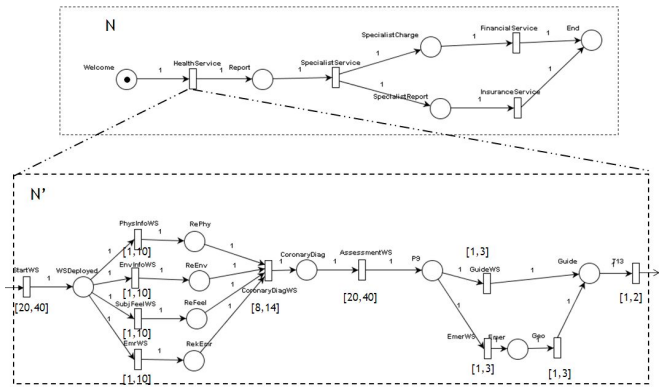

Fig. 9. A healthcare process modeled by Hierarchical Timed Predicate Petri nets (HeathService level)

The verification of a healthcare system model is the key before deploying them into operation, especially when a mission critical service does not tolerate run-time failure due to design errors. The healthcare process that is represented using our hierarchical timed predicate Petri-nets model can be analyzed using a variety of analysis techniques:

- Reachability: the outcome of this analysis method is the acknowledgement of the designer about possibility of healthcare process to achieve the desired results or not.
- Boundedness: the healthcare process design is indicating an error each time the number of token in a given place is neither 0 nor 1.
- Timing constraints: this method allows the designer to find whether a healthcare process is schedulable under the consideration of timing constraints without actual execution based on its specification [8]. By comparing the time span with firing during, time consistency property can be verified.

## 6 CONCLUSION

In this paper, we presented a novel architecture of healthcare services platform. Using hierarchical timed Petri-nets as the basis for modeling, we demonstrate that complexity of healthcare process model can be reduced and we are able to add conveniently timing constraints specification to healthcare process model.

Further research is needed for sure. Firstly, the customization layer should be an adapter module to capitalize from the past of citizen activity and formalize the evaluation process. Secondly, we will extend the Petri-Net based model to support dynamic QoS properties for future work.

## REFERENCES

[1] S. Rey, and H. Sarrazin, "How telecoms can get more from Internet Protocol," *McKinsey on IT*, http://www.mckinsey.de/downloads/publikation/mck_on_bt/2006/MoIT10_IP%20transformation.pdf. 2006.
[2] S. Mtibaa, and M. Tagina, "An Automated Petri-Net Based Approach for Change Management in Distributed Telemedicine Environment," *Journal of Telecommunications*, vol. 15, no. 1, pp. 1-9, 2012.
[3] L. Chen, "Integrating Cloud Computing Services Using Enterprise Service Bus (ESB)," *Business and Management Research*, vol. 1, no. 1, pp. 26-31, 2012.
[4] A. Azees, "Challenges and Performance Enhancement in Could Computing," *IRACST - International Journal of Computer Science and Information Technology & Security (IJCSITS)*, vol.1, no. 1, pp. 49-54, 2011.
[5] P. Wang, Z. Ding, C. Jiang, and M. Zhou, "Web Service Composition Techniques in a Health Care Service Platform," *IEEE International Conference on Web Services*, pp. 355-362, 2011.
[6] V. .s. Kondapall, J. Raju, K. R. Rao, and K. Tanuja ," Patient-Centric-Integrated EHA Of Physician Practice Portal Through Cloud Computing Technology," *International Journal of Computer Science & Communication Networks*, vol. 2, no. 1, pp. 172-18, 2012.
[7] E. Badouel, M. Llorens, and J. Olivier, "Modeling Concurrent Systems: Reconfigurable Nets," *Proc. Int. Conf. on Parallel and Distributed Processing Techniques and Applications*, pp. 1568-1574, 2003.
[8] G. Dai, R. Liu, C. Zhao, and C. Hu, "Timing Constraints Specification and Verification for Web Service Composition", *Asia-Pacific Services Computing Conference*, pp. 315 – 322, 2008.

**Sabri Mtibaa** is currently a Ph.D. student in the National School for Computer Sciences of Tunis, Tunisia (ENSI). He received the master degree from High School of Communication of Tunis, University of Carthage, Tunisia (Sup'Com) in 2008. His current research interest includes web service composition using Petri nets as well as system verification and QoS aware.

**Moncef Tagina** is professor of Computer Science at the National School for Computer Sciences of Tunis, Tunisia (ENSI). He received the Ph.D. in Industrial Computer Science from Central School of Lille, France, in 1995. He heads research activities at LI3 Laboratory in Tunisia (Laboratoire d'Ingénierie Informatique Intelligente) on Metaheuristics, Diagnostic, Production, Scheduling and Robotics.